# APPLICATION OF FREQUENCY MAP ANALYSIS TO BEAM-BEAM EFFECTS STUDY IN CRAB WAIST COLLISION SCHEME

D.Shatilov, E.Levichev, E.Simonov
*BINP, Novosibirsk 630090, Russia*

*Abstract*

We applied Frequency Map Analysis (FMA) – a method that is widely used to explore dynamics of Hamiltonian systems – to beam-beam effects study. The method turned out to be rather informative and illustrative in the case of a novel Crab Waist collision approach, when "crab" focusing of colliding beams results in significant suppression of betatron coupling resonances. Application of FMA provides visible information about all working resonances, their widths and locations in the planes of betatron tunes and betatron amplitudes, so the process of resonances suppression due to the beams crabbing is clearly seen.

## 1. INTRODUCTION

Crab Waist (CW) collision scheme was proposed in [1] to enhance the luminosity of electron-positron colliders. The idea of CW thoroughly examined in [2,3] is, in brief, as follows. Two bunches with small transverse sizes (low emittance beams are essential) intersect at large Piwinski angle, so the length of the overlap area is much smaller than the bunch length. In this case the vertical beta function at IP can be squeezed to the length of the intersection area (sub-millimeter range) without incurring in the hour-glass effect, so the luminosity enhances substantially. On the other hand, betatron coupling beam-beam resonances are strongly excited in such a scheme, thus limiting the maximum achievable tune shift $\xi_y$. This drawback can be overcome by introducing the CW transformation which is realized by two sextupole magnets placed in phase with the IP in the horizontal plane and at $(2n+1)\cdot\pi/2$ in the vertical plane as it is shown in Fig. 1 [3].

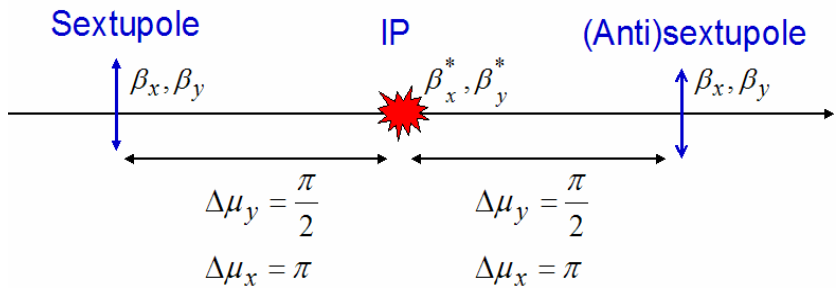

Fig. 1 Crab Waist sextupoles location.

The sextupole strength estimated from optical consideration should be:

$$\left(\frac{B''l}{B\rho}\right)_{CW} = \frac{1}{\theta}\frac{1}{\beta_y^*\beta_y}\sqrt{\frac{\beta_x^*}{\beta_x}}, \qquad (1)$$

where $\theta$ is the full crossing angle and the other parameters are explained in Fig. 1. The sextupoles focus particles locally in such a way that the vertical waist (minimum of $\beta_y$) rotates and adjusts along the axis of the opposite beam, as it is illustrated in Fig. 2. The CW transformation provides effective suppression of betatron coupling resonances (together with their synchro-betatron satellites), thus increasing the $\xi_y$ limit by a factor of about three. Firstly it was predicted by simulations and then observed experimentally at DAΦNE Φ-factory [10].

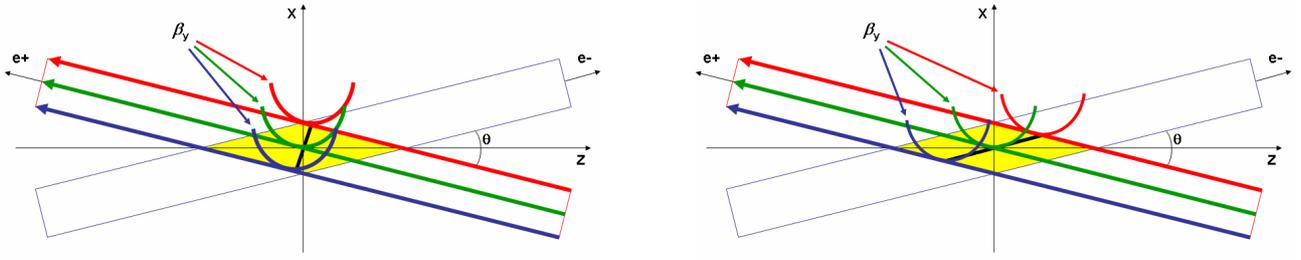

Fig. 2 Vertical beta function at IP without (on the left) and with the CW transformation.

Frequency Map Analysis [4,5] turned out to be very useful for CW investigations, as it provides visible information about all working resonances, their widths and locations in the planes of betatron tunes and betatron amplitudes, so the process of resonances suppression due to the beams crabbing is clearly seen. We applied this technique to see how CW works at DAΦNE Φ-factory, where a very good agreement between simulations and experimental data has been obtained [11]. Then we performed similar studies for SuperB project [14]. Many interesting observations were made which considerably enriched our understanding of beam dynamics in CW scheme.

## 2. FREQUENCY MAP ANALYSIS

At the phase plot of a Hamiltonian system one can see a complicated mixture of periodic, quasiperiodic and chaotic trajectories arranged in stable and unstable areas. Analysis of these trajectories and distinction between regular (periodic or quasiperiodic) and chaotic ones may give useful information on the motion features. One of the possible techniques providing such info on every particular trajectory is FMA proposed by J.Laskar [4,5]. In accelerator community this method is used mainly for dynamic aperture study, see, for example,[6].

Since the FMA technique is commonly used, we only briefly mention some peculiarities of its implementation in our studies. For any given initial coordinates we track a particle for 2024 turns and use rectangular window of 1024 turns to determine the betatron tunes. Then the window is shifted 20 times by 50 turns, so we get 21 set of numbers (see Fig. 3b, bold dots). The diffusion index is calculated as $\text{Log}_{10}(\sigma_\nu)$, where $\sigma_\nu$ is the r.m.s. spread of tunes.

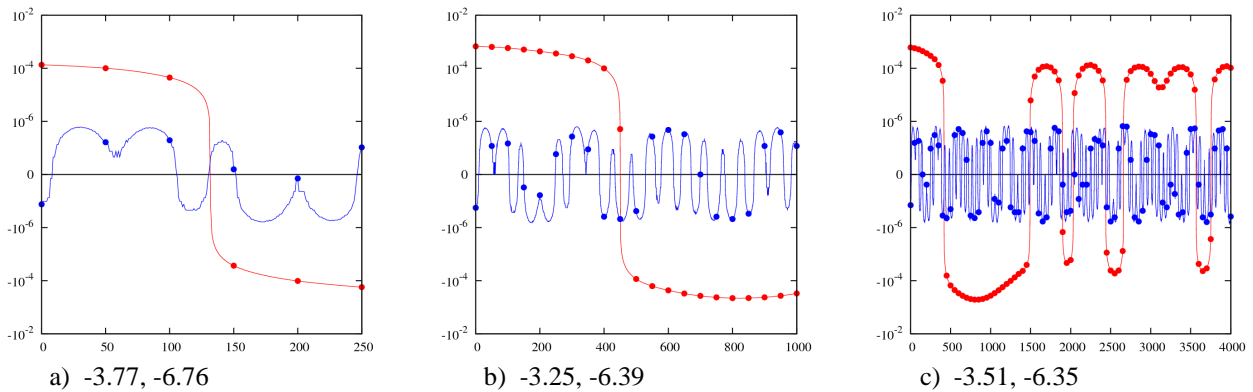

a)  -3.77, -6.76
b)  -3.25, -6.39
c)  -3.51, -6.35

Fig. 3 Betatron tune deviation versus window shift for chaotic (red) and regular (blue) trajectories in different time scales. The diffusion indexes calculated from the bold dots are shown at the bottom.

An example of more detailed time-frequency dependences for two different trajectories is shown in Fig. 3, where the horizontal axis represents the window shift in turns (bold dots correspond to 50-turn intervals) and the vertical one – the tune deviation from the average in logarithmic scale. As we see, the calculated diffusion indexes depend rather weakly on the time scale.

## 3. BEAM-BEAM INTERACTION IN DAΦNE WITH CRAB WAIST

The DAΦNE Φ-factory was upgraded in the second half of 2007 in order to increase luminosity and test the Crab Waist idea [7,8]. As a result the peak luminosity was boosted by a factor of about three. The gain could be even larger, but it was limited by the crab sextupoles strength and the effects disturbing positron beam at large currents (e.g. electron cloud instability) [9,10]. In order to investigate the innovative CW collision scheme and find its abilities and limitations due to the beam-beam effects, a "weak-strong" experiment was carried out in May 2009; the results will be published soon by the DAΦNE Team. In our studies we used the machine parameters corresponding to the best luminosity achieved in this experiment, see the Table.

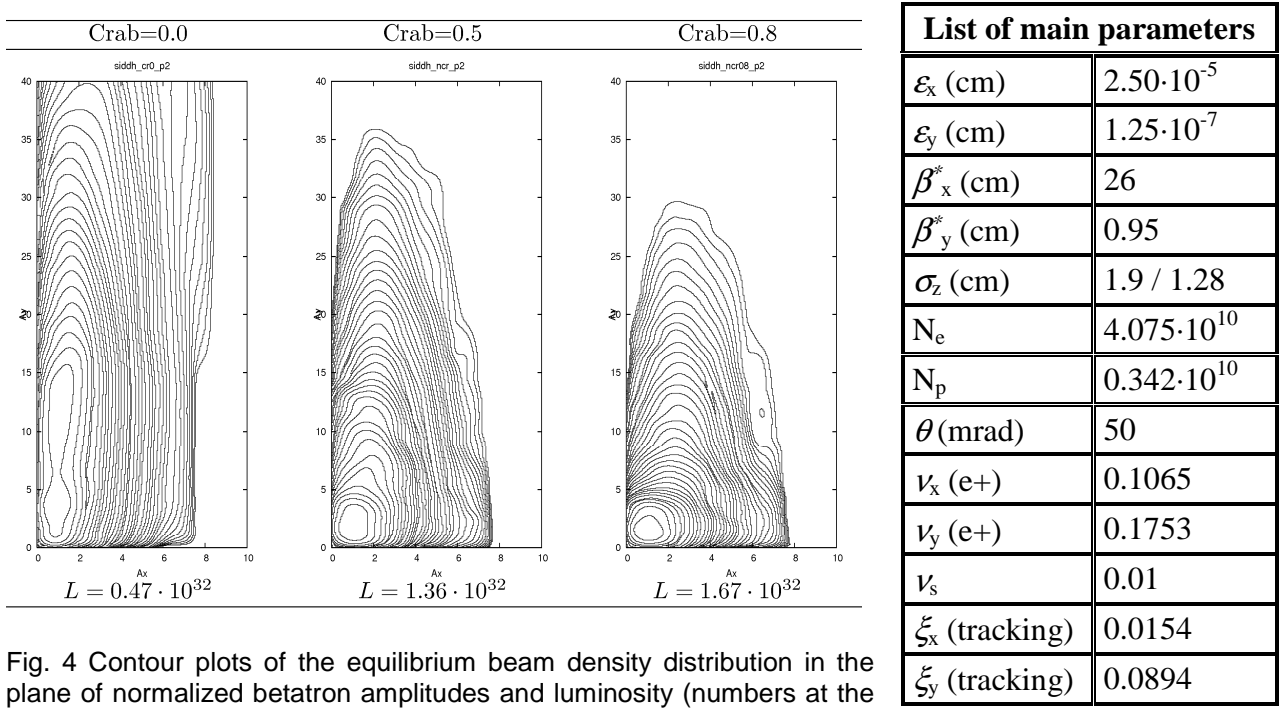

| List of main parameters | |
|---|---|
| $\varepsilon_x$ (cm) | $2.50 \cdot 10^{-5}$ |
| $\varepsilon_y$ (cm) | $1.25 \cdot 10^{-7}$ |
| $\beta^*_x$ (cm) | 26 |
| $\beta^*_y$ (cm) | 0.95 |
| $\sigma_z$ (cm) | 1.9 / 1.28 |
| $N_e$ | $4.075 \cdot 10^{10}$ |
| $N_p$ | $0.342 \cdot 10^{10}$ |
| $\theta$ (mrad) | 50 |
| $\nu_x$ (e+) | 0.1065 |
| $\nu_y$ (e+) | 0.1753 |
| $\nu_s$ | 0.01 |
| $\xi_x$ (tracking) | 0.0154 |
| $\xi_y$ (tracking) | 0.0894 |

Fig. 4 Contour plots of the equilibrium beam density distribution in the plane of normalized betatron amplitudes and luminosity (numbers at the bottom) versus the crab sextupoles strength.

Simulation results of the equilibrium density distribution for different crabbings are shown in Fig. 4. Actually, the waist rotation in DAΦNE was limited by the strength of crab sextupoles, so in reality it can reach only 0.5 of the nominal value, while the maximum luminosity is expected at the crab value of 0.8. On the other hand, without crabbing, not only luminosity drops down significantly but the "weak" bunch dies due to the long tails induced by beam-beam interactions. It means that for the given electron current, we can compare the simulation results with the experimental data only for crab=0.5; and the obtained agreement was almost perfect. This gave us more confidence in the tracking code which then was used in our FMA studies.

Usually we build the FMA plots in two planes: betatron tunes and normalized betatron amplitudes. Although the definition of normalized amplitudes for nonlinear trajectories is rather questionable, the information provided by FMA in the amplitude space is very useful and interesting, so in order to get these plots we allowed some simplifications. First of all, we use the linear unperturbed (without beam-beam) transport matrix to calculate the normalized betatron amplitudes from the physical coordinates. Of course, such amplitudes will not conserve along a trajectory due to nonlinearities, but this is not important for our purposes as we plot the diffusion indexes versus the *initial* betatron amplitudes. Secondly, if we examine different betatron phases for the given amplitude we can get rather different diffusion indexes for these trajectories, so the picture of resonances will be blurred. To avoid this side effect, we examine only *one* betatron phase for the plots in the amplitude plane, while for the plots in the tune plane many different phases are considered in order to get a more complete picture.

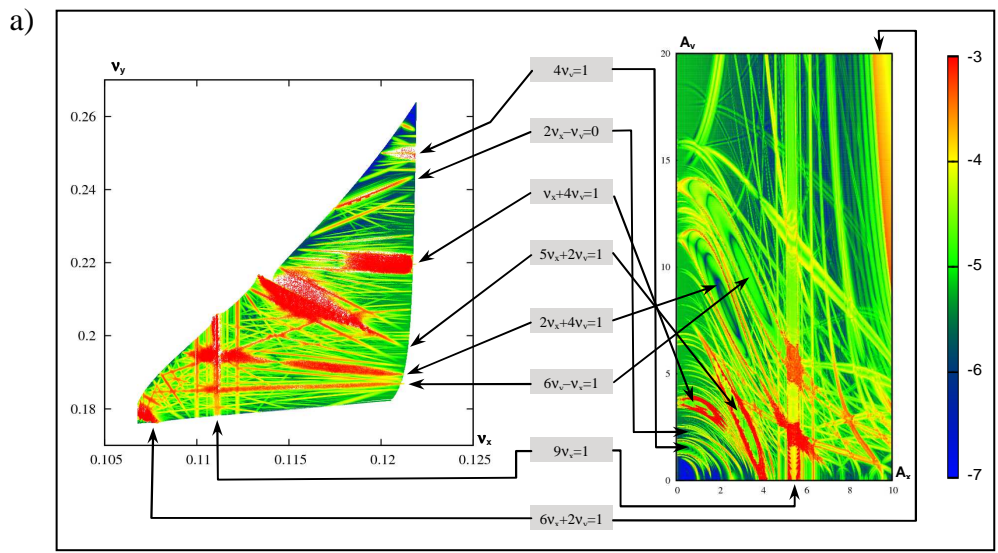

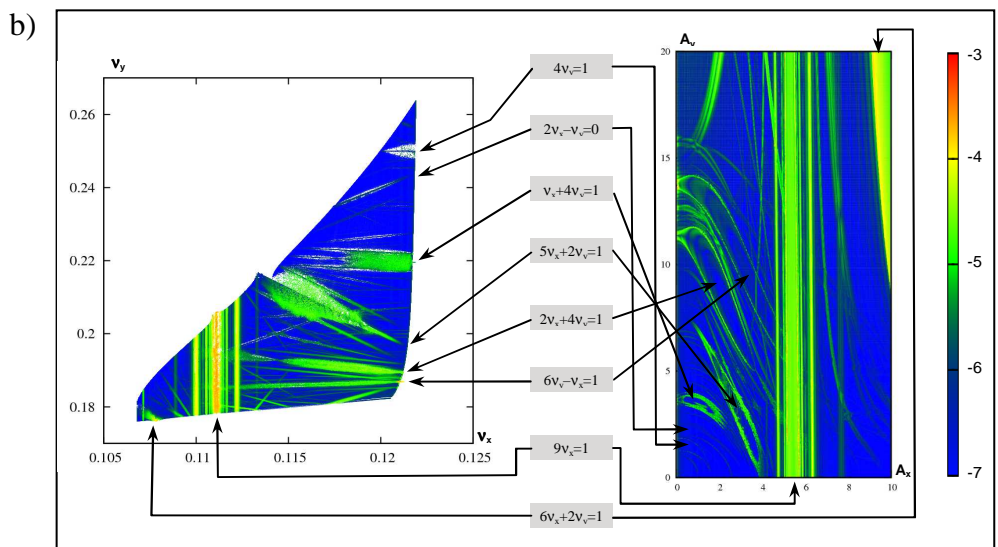

Fig. 5 Beam-beam resonances in the tune and amplitude planes for Crab=0.4. Correspondence between color and diffusion index is shown in the color palette on the right. In the case b) the diffusion index was calculated from the horizontal tune spread only.

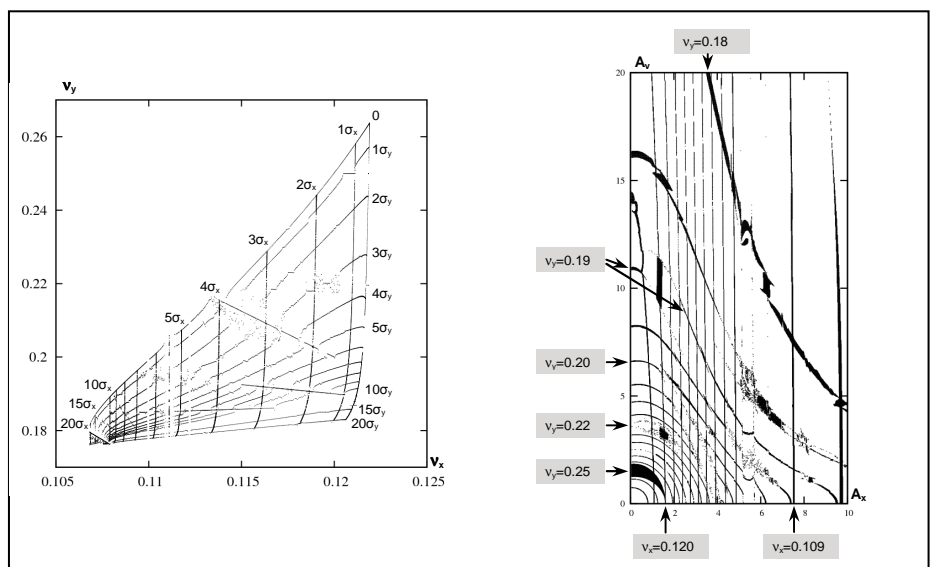

Fig. 6 Transformation between betatron tunes and betatron amplitudes.

An example of the output data is presented in Fig. 5, where the FMA plots for DAΦNE with Crab=0.4 are shown. Normally we define the trajectory's diffusion index as the maximum of two values calculated from the horizontal and the vertical tune spreads, see Fig. 5a. In order to estimate the relative contributions of these two components, we also built the plots using only the horizontal tune analysis, see Fig. 5b. In the latter case, the "spectrum" is distinctly shifted to the blue side in accordance with the well-known feature of flat beams, where the perturbations occur mainly in the vertical plane.

Apparently, many beam-beam resonances including the high-order ones can be clearly seen and identified on both planes. Among them one can distinguish synchro-betatron resonances of different orders – in the plane of tunes they are parallel to the generative betatron ones. It is interesting that due to the crossing angle, the opposite beam excites small synchrotron oscillations for the particles with zero initial synchrotron amplitudes. Though this excitement is not significant, within one percent of sigma, it is enough to detect synchro-betatron resonances. But in order to make them more observable we set the initial synchrotron amplitude to 0.1 sigma.

Note how the strong resonances look like. In the amplitude plane their widths can be recognized by the red contours corresponding to the separatrix, while in the center of resonance the motion is more regular (green and blue colors). It is interesting that in the plane of tunes, strong resonances can be surrounded by specific white areas. Indeed, all trajectories within a resonance island have the same frequencies satisfying the resonance equation, so in the plane of tunes they are attracted to the resonance line, thus forming a gap around it. The red points within white areas come from the near-separatrix region, where the tune-amplitude dependence has the largest spread.

The transformations between betatron tunes and betatron amplitudes are also affected by strong resonances, see Fig. 6. The ordinary lines of constant amplitudes in the footprint (see the plane of tunes) are noticeably disturbed by the trajectories located within the resonance islands and thereby attracted to the resonance lines. As for the reverse transformation, some explanations are required concerning the algorithm of building the lines of constant tunes in the amplitude plane. Indeed, it is easy to assign any given amplitudes to a test particle, but it is impossible to assign tunes as they remain unknown until the actual trajectory is analyzed. Normally, to produce the FMA plots we track many particles and scan their initial normalized amplitudes (both horizontal and vertical) with a step of 0.02. Then, in order to build the lines of constant tunes, we define some allowance (or gap) around the given values and plot all the points which fall within it. Of course, this approach has a number of side effects: the obtained "lines" are rather thick and shaky, sometimes even non-continuous. Besides, when they cross strong resonances under small angles, the specific thickenings can be formed – see the right part of Fig. 6.

Good visualization of different resonances and easy estimation of their widths and strengths, provided by the FMA technique, make this method very useful for the Crab Waist investigations. In order to study how the resonances are suppressed by the CW transformation we performed a scan of crab value in the range of 0 to 1 with a step of 0.1, see Figs. 7 and 8. One of important features facilitating the comparison is that the location of resonances in the tune and amplitude planes is almost not affected. Yet note how the area occupied by the footprint shrinks when the resonances are suppressed.

Actually we need only the weak beam crabbing to suppress the resonances. But in our simulations, to comply with reality, we applied the same CW transformation to both beams. Crabbing of the strong beam makes its distribution essentially non-Gaussian, so the well known Bassetti-Erskine formulae become non-applicable for the beam-beam kicks calculations. To solve the problem, a new feature was implemented in the LIFETRAC tracking code, which allowed calculating the kicks from arbitrary beam distributions using the prepared-in-advance *grid* files. Though the effects of the strong beam crabbing are not significant, some of them have to be mentioned. Firstly, the optimum crab value slightly increases: from 0.6 to about 0.8. That conforms to the strong-strong simulations by K.Ohmi and analytical estimations by M.Zobov [12]. Secondly, the geometrical luminosity increases by a few percents, in accordance with the analytical estimations [13]. And thirdly, the actual tune shifts are also affected: note how the footprint height increases with the beams crabbing.

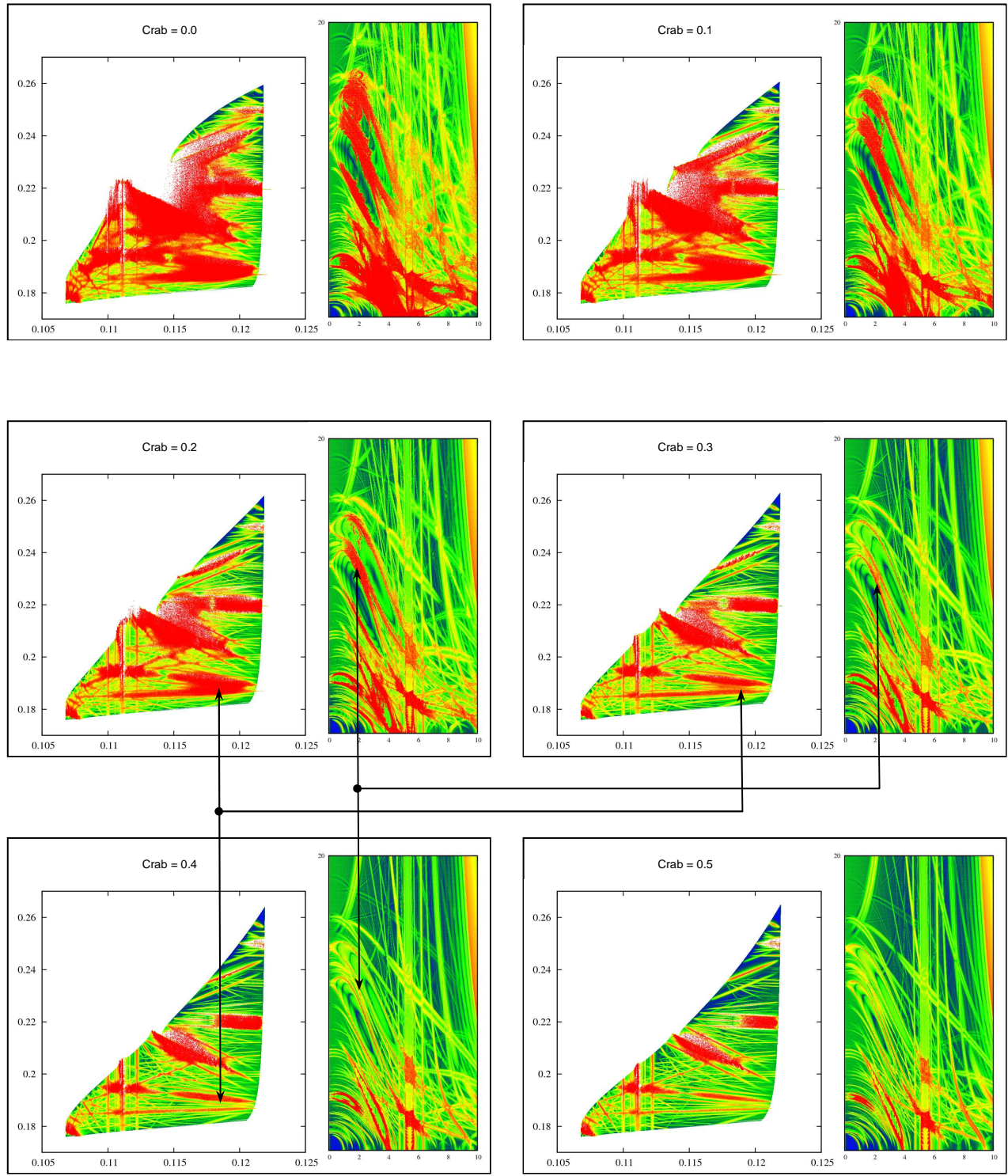

Fig. 7 Beam-beam resonances versus the crab value for DAΦNE (part 1).

Using the FMA technique we also have found a good illustration on the well-known Chirikov's criterion of stochasticity. Pay attention to the resonances $2\nu_x + 4\nu_y = 1$ and $6\nu_y - \nu_x = 1$ – they are strong enough, close to each other, and isolated from the other strong resonances in some area of their locations. When the beam crabbing decreases from 0.4 to 0.2, their widths increase and start to overlap, thus creating a stochastic layer in the overlapping region, see the areas indicated by arrows in Fig. 7. The effect is clearly seen on both the tune and the amplitude planes, but the latter one seems to be more relevant as it allows a better recognizing detection of the resonance widths.

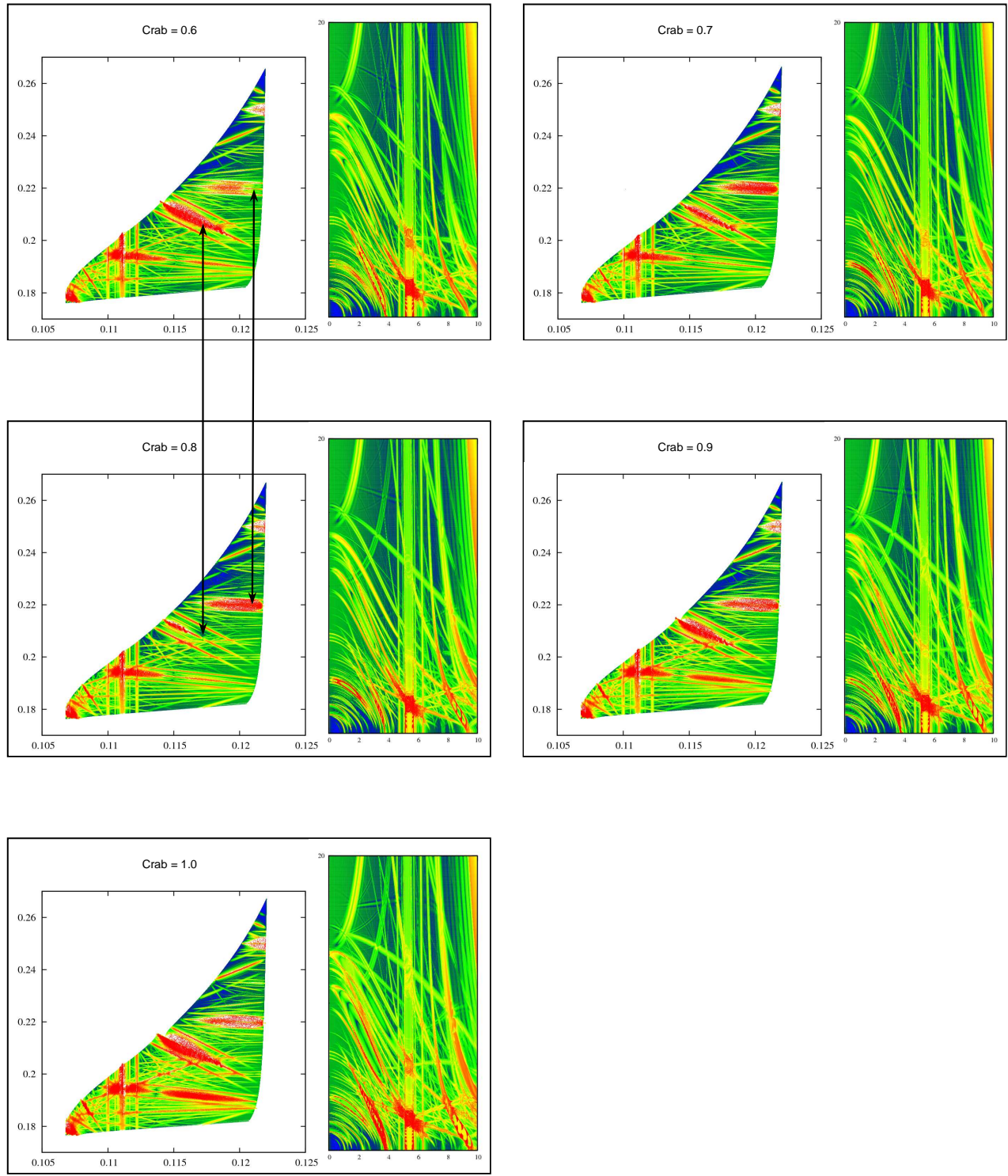

Fig. 8 Beam-beam resonances versus the crab value for DAΦNE (part 2).

As it was explained in [2], the optimum angle of waist rotation usually is less than the nominal value. The reason is that, basically, there are two mechanisms exciting resonances due to the horizontal betatron oscillations. The first and the most important one is the vertical betatron phase modulation, which is suppressed by the nominal CW transformation (crab=1). But amplitude modulation of the vertical kick can be only partially suppressed, and here the optimum crab < 1. When considering both mechanisms, obviously, the optimum must be < 1, but the actual value can depend on particular resonances, e.g. take a look at the resonances $\nu_x + 4\nu_y = 1$ and $5\nu_x + 2\nu_y = 1$ pointed out by arrows in Fig. 8. Evidently, the optimum crab value is 0.6 for the first one and 0.8 for the second. As for the whole picture, the optimum value lies somewhere between 0.7 and 0.8.

## 4. BEAM-BEAM SIMULATIONS FOR SUPERB

In the current version of SuperB design [16], there is an asymmetry between HER and LER lattices: emittances and beta functions are different, though the vertical beam sizes at IP are the same. Such asymmetry noticeably affects the beam dynamics. In particular, the hour-glass effect and the vertical betatron phase modulation (without CW) are much more pronounced in LER [15]. As a result, the beam-beam effects become much stronger for LER regardless of the fact that the "nominal" tune shifts are the same for both rings. In addition, the optimum crabbings become different too: 0.8 for HER and about 1.0 for LER. On the other hand, the designed beam-beam tune shift $\xi_y$ is far below the limit, which is about 0.2 for the SuperB parameters with CW, so we have a rather large margin of safety. This widens our possibilities in choosing the working point, but in any case there is a need in beam-beam simulations, especially for LER which is the weak point. Since the SuperB lattice is not finalized yet and additional Dynamic Aperture optimizations are required, in our simulations we used a linear lattice with the given parameters (see the Table), where the crab sextupoles and the opposite beam were the only nonlinear elements.

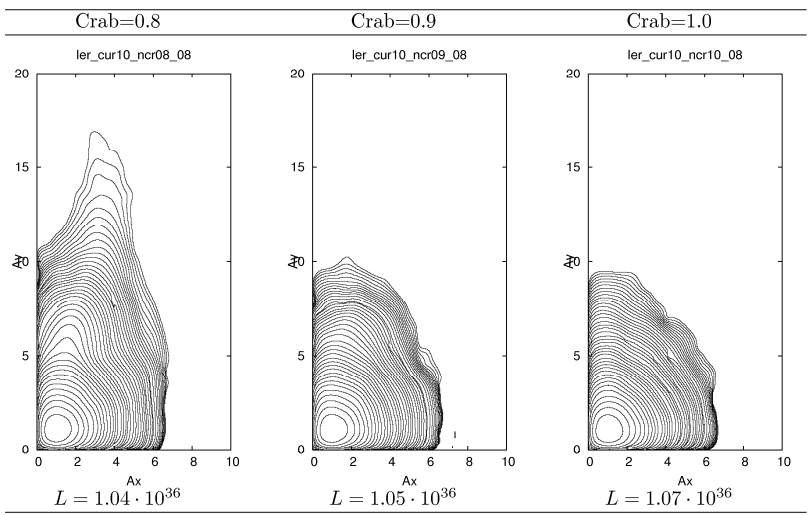

Fig. 9 Contour plots of the equilibrium beam density distribution for SuperB LER and luminosity (numbers at the bottom) versus the crab sextupoles strength.

| List of parameters LER / HER | |
|---|---|
| E (GeV) | 4.18 / 6.7 |
| $\varepsilon_x$ (cm) | $(2.56 / 1.6) \cdot 10^{-7}$ |
| $\varepsilon_y$ (cm) | $(6.4 / 4.0) \cdot 10^{-10}$ |
| $\beta^*_x$ (cm) | 3.2 / 2.0 |
| $\beta^*_y$ (cm) | 0.02 / 0.032 |
| $\sigma_z$ (cm) | 0.5 |
| $N_p$ | $5.74 \cdot 10^{10}$ |
| $\theta$ (mrad) | 60 |
| $\nu_x / \nu_y / \nu_s$ | 0.542 / 0.58 / 0.01 |
| $\xi_x$ (tracking) | 0.0047 |
| $\xi_y$ (tracking) | 0.1063 |

It is worth mentioning that the main profit of Crab Waist comes from the fact that it allows an essential increase of the beam-beam tune shift. In other words the effect of CW becomes valuable only for large $\xi_y$ (high bunch currents) and it decreases when $\xi_y$ is getting relatively small. The latter was confirmed once again by our simulations, see Fig. 9. Though the optimum there can be determined, both the luminosity and the beam tails remain almost the same for crab value in the range of 0.8 to 1.0, in contrast to higher bunch currents ($\xi_y$ about 0.17÷0.20) where dependence on crabbing becomes much stronger. It is interesting that the FMA technique turned out to be very sensitive to the fine tuning of parameters. So, we processed exactly the same three cases shown in Fig. 9, but in FMA plots the differences look much more pronounced, see Fig. 10.

Note also that a number of interesting observations can be made from the footprints (see the plane of tunes in Fig. 10) even without FMA. First of all, pay attention to the footprint shape – how it differs from the "classical" one and how strongly it depends on the crabbing. One more surprise is connected with the actual horizontal tune spread in the beams. As it is seen in Fig. 9, the equilibrium beam distribution is located well within 6 $\sigma_x$ horizontally. Looking at the footprint for crab=0.8, where the grid of betatron amplitudes (5, 10, 15 and 20 sigma) is shown, we may conclude that the actual spread of $\nu_x$ is not greater than 0.0003. That is about 1/15 of the horizontal tune shift $\xi_x$ which is also very small itself. This can be considered as one more positive feature of colliding scheme with large Piwinski angle.

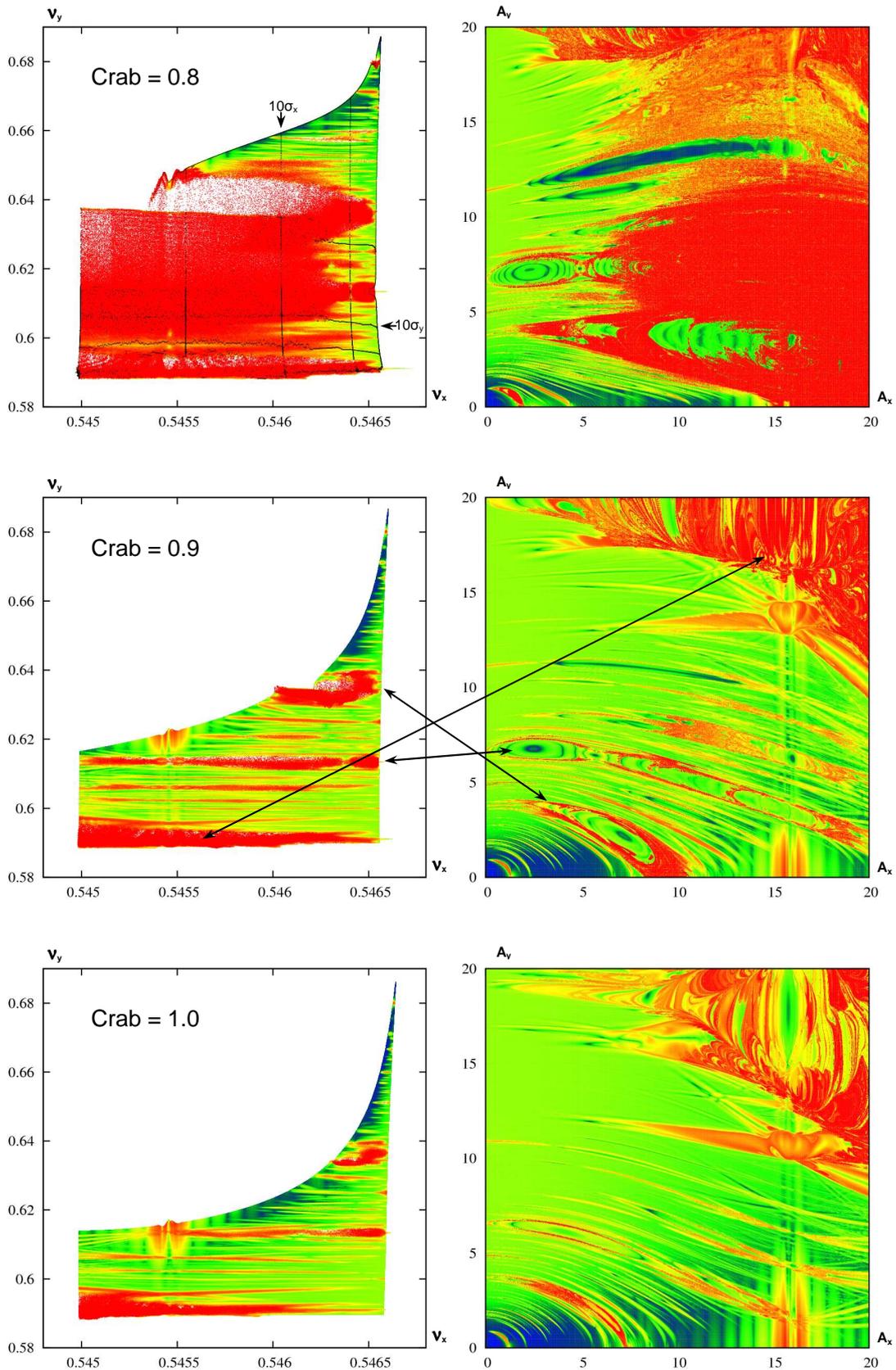

Fig. 10 Beam-beam resonances versus the crab value for SuperB LER.

In order to make the FMA plots more informative and allow more resonances to be identified we enlarged the plotting area to 20 sigma in both directions, while the actual beam density occupies only a small part (in the plane of tunes – close to the peak). As we see, the main differences are located at larger amplitudes, that is why they are not seen in Fig. 9.

## 5. FMA VERSUS OLD-STYLE TRACKING

First of all we have to remember that both methods use the same tracking code, so they are inter-consistent and cannot check each other. The main differences are connected with the simulation of radiation damping and noises. The *old-style* tracking produces the luminosity and equilibrium density distribution (e.g. see Figs. 4 and 9), for that damping and noises must be taken into account. On the other hand, identification of resonances in the density contour plots is rather difficult: only strong isolated resonances sometimes can be detected.

On the contrary, tracking for FMA must be without noise and damping; this results in a very high resolution of resonances: even high order resonances can be clearly identified. Therefore, FMA can be very useful for investigating particular resonances, their strengths, widths, locations, and the influence of different conditions. On the other hand, tracking for FMA is more time-consuming (more particle-turns are required), and it cannot give the numbers for luminosity and density in the beam tails (lifetime).

Thereby we come to the conclusion that these two techniques are mutually complementary. Together they give more complete understanding, in some sense *stereoscopic view* on the nonlinear beam dynamics. We should use both of them, as they answer different questions.

## 6. CONCLUSION

We found the FMA technique to be very useful for beam-beam interaction studies. Demonstration of how the Crab Waist works is clear and impressive. The capabilities of investigating every particular resonance, provided by FMA, can be very helpful for a better understanding of various mechanisms affecting the nonlinear beam dynamics.

## ACKNOWLEDGEMENTS

The authors greatly appreciate the long-lasting fruitful collaboration with the DAFNE team, whose remarkable achievements inspired us to perform these studies. All simulations were performed on the computing cluster of Novosibirsk State Technical University, Department of Computer Engineering. The authors would like to thank the department's staff for the provided resources and technical support.